\documentclass[%
onecolum,
 aip,
 jcp,%
 amsmath,amssymb,
reprint,%
]{revtex4-1}
\usepackage[ansinew]{inputenc}
\usepackage{graphicx}
\usepackage{dcolumn}
\usepackage{bm}
\usepackage{amssymb,amsmath,latexsym,amsfonts}
\usepackage{amsthm,subfigure}
\usepackage{amsbsy}
\usepackage[mathscr]{eucal}
\usepackage[dvipdf]{epsfig}
\usepackage{color}

\begin{document}

\title{Minimal model of an active solid deviates from equilibrium mechanics}
\author{Mario Sandoval}
\email{sem@xanum.uam.mx}
\affiliation{
 Department of Physics, Complex Systems, Universidad Autonoma Metropolitana-Iztapalapa, 
 Mexico City 09340, Mexico.
}

\begin{abstract}
In this work, the mechanical response  of  an one-dimensional active solid --defined as a network of  active stochastic particles interacting by nonlinear  hard springs-- subject to an external deformation force, 
is numerically studied and rationalized with a minimal model. It is found that an active solid made of linear springs and subject to an external stress,
presents an average deformation  which is independent of the system's activity. However, when the  active solid is made of nonlinear hard springs, the solid's average deformation decreases with respect to a passive system under the same conditions, and as a function of activity and rotational noise in the system. The latter result may shed light on new ways to creating an active metamaterial, which could tune its stiffness by moving either its activity or rotational noise.
\end{abstract}

\date{\today}
\maketitle%

\section{INTRODUCTION}

{The materials science community \cite{Siga,Siga2,Kush,science,Barry} is already designing and characterizing metamaterials or metastructures defined as synthetic  materials with deliberately designed structures. These metastructures are found to present exotic  behavior such as the ability to prevent wave propagation through them \cite{Siga,Siga2,science} (the so-called bandgap), negative Young's modulus \cite{massY},  a negative index of refraction \cite{index}, or a negative effective mass  density \cite{massY}. After the inclusion of nonlinearities, those metastructures are seen to present  asymmetric wave traveling phenomenon \cite{asy,Zivi,Fiore} and the development of gap solitons \cite{gap,enve}. Quite recently, and by exploiting 3D printing techniques researchers have designed metamaterials with programmable response to uniaxial compression \cite{hecke}, a negative Poisson's ratio \cite{Bertoldi} or even materials with a  nearly constant stiffness at very low densities \cite{Zheng}. Metamaterials are also being used for vibration suppression and  energy harvesting \cite{Huenergy,Li,Tol,Moha,asme,Zivi,Fiore}. }
 
{ In addition, the physics community has also spotted a different way to render a metamaterial a different behavior. The key is to take this material out of equilibrium by imposing a propulsion or active force to each of its building components \cite{Ferrante,Menzel,Lowenmaterial,Kaiser,Peruani}, thus creating an active metamaterial. Explicit recent works on elucidating a change of a material's behavior due to its non-equilibrium state are: Souslov \cite{vicenzy} {\it et al.} who built a lattice made of annular shapes filled with an active fluid. Using numerical simulations, they observed the propagation of  unidirectional sound waves through the edges of their system domain. Wang  \cite{breaking} {\it et al.}  and Trainiti \cite{Trainiti} {\it et al.}  proposed a one-dimensional crystal whose elastic interactions are time-dependent. This time-dependence breaks time-reversal symmetry, thus creating a directional bandgap in the dispersion equation.
Brandenbourger \cite{Martin} {\it et al.} used a chain of springs with  broken spatial-symmetry, that together with actuation and local sensing, observed the emergence of asymmetric standing waves with unidirectional amplification. As it can be noticed, scientists have already spotted self-propulsion as a means of creating novel active metamaterials.}

\begin{figure}
\includegraphics [width=8.2cm]{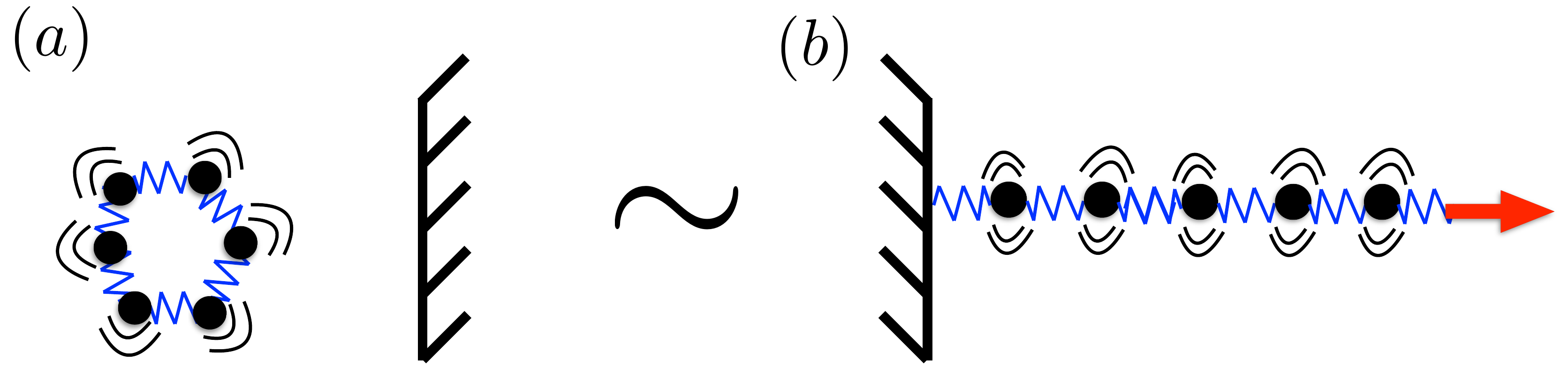}
\caption{{(a) A chain of active particles with hexagonal shape about to collide against a wall \cite{Martin2}. This chain was seen to rebound differently when its activity was turned off or on.
(b) By only keeping the main ingredients from (a), we build the active structure under stress (red arrow) shown in (b). Thus, this work proposes to give an analytical understanding of possible activity effects on the deformation of this simpler active metastructure.}}
\label{s0}
\end{figure}

{Motivated by the latter efforts on rendering a material new properties and by recent active materials experimental observations, such as the peculiar bouncing over a wall of a closed chain of active robots interacting by mobile active arms \cite{Martin2} (see cartoon in Fig. \ref{s0}(a) for illustration), and its bouncing dependence on
activity; or a network  of active robots interacting by
means of springs and generating  a synchronized dynamics \cite{Paul}; or a  larger structure made of several
closed active chains as its main unit (see Fig. \ref{s0}(b) for a possible realization), and observed to deform
differently when  activity in the structure was turned on \cite{Martin2}; 
I report on numerical and analytical
results obtained after analyzing an active solid
made of particles interacting by nonlinear hard springs described by constants $k$ and $\mu$. This simple one-dimensional model retains the main ingredients (interaction and self-propulsion) of the previous experiments \cite{Martin2,Paul}, and can be used to evaluate the effect of those ingredients on the active solid's mechanical response to a deformation force (stress). Briefly, it is found that the latter active solid deforms less on average, as the activity in the
system grows. The results of this model may help to understand the mentioned experimental
observations and could be generalized to higher dimensions. 
It is interesting to mention that when seeking a model able to provide insight on possible self-propulsive effects on a material's response, it was observed that a linear interaction between building blocks was not able to do so; instead, nonlinear interactions \cite{FPUO} proved to be a valuable tool to accomplish that.}  {It is worth mentioning that a nonlinear mechanical response of  ephitelial monolayers to stress, has been experimentally spotted \cite{Xavier,charras}. Numerical and analytical efforts highlighting the nonlinear response of a tissue-model have also been reported \cite{Lisa,Huang}. Moreover, internal active forces as the responsible for stiffening a biopolymer network \cite{Dogic}, as well as the responsible for  a phase transition from a ductile to a brittle state in ephitelial tissue of {\it Trichoplax Adhaerens}, and for a bistable fission process (asexual reproduction) have already been experimentally identified \cite{Vivek}.  Therefore, this active nonlinear model  extended to higher dimensions may also be suitable for analyzing the dynamics of living tissues.} { Note that although active forces have been already identified to play an important role on the dynamics of artificial active metamaterials and ephitelial tissue, a minimal model able to encompass elasticity and active forces, and to systematically produce a change in the system's elastic or dynamic properties --after tuning certain parameters-- has not been proposed. This  study is an attempt to fill this gap.}

{
The present study is organized as follows: Section \ref{model} introduces the
model. Section \ref{Deformation} studies the deformation --due to an external force-- of  active solids made of self-propelled
particles either linearly or nonlinearly interacting and as a function of
activity in the system, rotational diffusivity, and strength of nonlinearities. An energy analysis of the system is carried out in Section \ref{Energetics}. Section \ref{expe} discusses available experiments on self-propelled robots and presents some orders of magnitude for relevant parameters such as propulsion force, dissipation coefficient, and angular velocity. Due to the similarity  of the orders of magnitude between the  used numerical parameters in Section  \ref{Deformation}  and the experiments, a possible experiment is proposed in this section.  A simple deterministic model to rationalize the present findings is introduced and discussed  in Section \ref{Ratio}. Conclusions and future directions
are offered in Sec. \ref{summa}, and two appendices complementing this work are also added. }

\section{Physical model}
\label{model}

\begin{figure}
\begin{center}
\includegraphics [width=8cm]{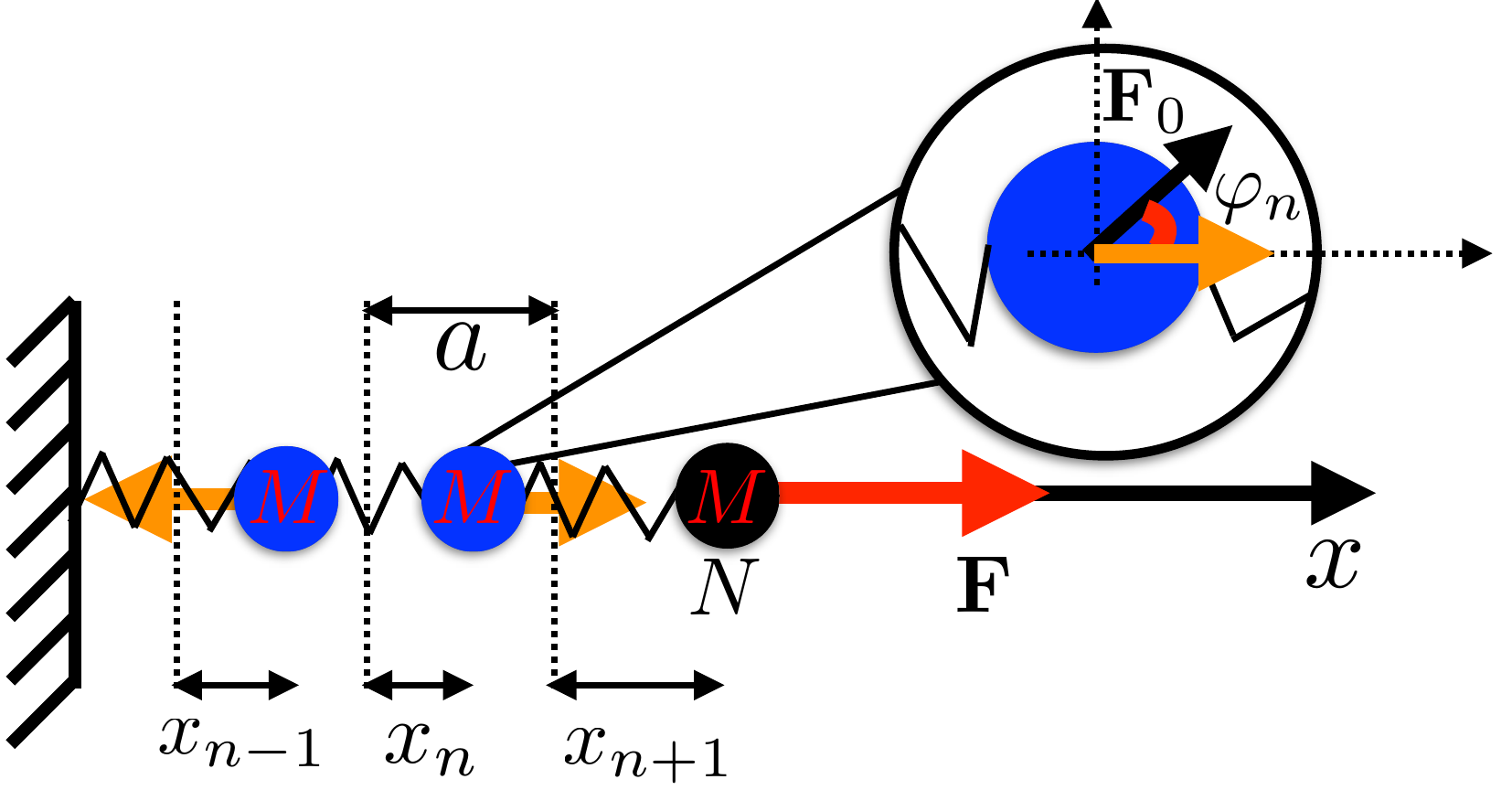}
\caption{{Schematic of the studied problem. One-dimensional active solid defined as a network of  $N-1$ active stochastic
particles (blue circles) of mass $M$, interacting by nonlinear springs. Note that the $N$ particle in this model is passive (black circle) and is subject to
a deformation constant force $\mathbf{F}$ (red arrow). In the picture, $N-1=2$. Due to the dimensionality of the system, their independent random propulsion
forces $\mathbf{F}_0$ have to be projected, that is,  $F_0\cos(\varphi_n(t))$. These projections are
indicated as orange arrows. The original equilibrium distance between particles is $a$, and no steric interactions are considered.}}
\label{diagram}
\end{center}
\end{figure}
Let us analyze a one-dimensional active solid defined as a network of $N-1$ active
stochastic particles of mass $M$, and interacting either by linear or nonlinear springs
characterized by constants $k$ and $\mu$; {where $k>0$ is the linear restitutive constant, and $\mu>0$ is the degree of nonlinearity}. In this system, the last
particle in the network, $N$, (see Fig. \ref{diagram}) is passive and its right hand side is not
subjected to a spring, but it is exposed to a deforming force (${\bf F}$) either stretching  or compressing the system. The active particles, originally separated   {by their equilibrium
distance $a$}, { are assumed to stochastically rotate along their azimuthal direction (with origin at their center) $\varphi
_{n}(t)$, and towards which a propulsion force ${\bf F}_{0}=(F_{0}\cos\varphi_n,F_{0}\sin\varphi_n)$ is acting. The latter is just an ABP (Active Brownian particles) model. Due to the one-dimensionality of this system, that may be accomplished by setting  smooth walls  in an experiment (see Section \ref{expe} and supplementary materials where an experiment is proposed), the propulsion forces will only be present through their projections along the $x$-axis, that  is, $F_{0}\cos(\varphi_n)$. See the inset of Fig. \ref{diagram} for further clarification.}  These active forces will be constantly driving the system out-of-equilibrium. { Note that dimensionality does not change the  effect of active forces on a system's effective diffusion. This has been proved by Lowen {\it et al.} \cite{nongau} who showed that in an one-dimensional active system, and even if the magnitude of the propulsion force varies according to $\cos(\varphi_n(t))$; it induces the same effective diffusion as in higher dimensions, and for which the magnitude of the propulsion force is kept constant. Thus we expect to have a similar behavior of our active solid at higher dimensions.}

The dynamics of each particle is described by its
translational velocity $v_{n}(t)$ and  position $x_{n}(t)$, with respect to
its equilibrium position. Therefore, the motion of the $n-th$ particle under a  {hard spring
symmetric potential,} $V(x)=\left( k/2\right) x^{2}+\left( 1/4\right) \mu
x^{4},$ and in dimensionless variables $\left( \widetilde{x}_{n},\widetilde{v%
}_{n},\widetilde{t}\right) $ reads  {
\begin{eqnarray}
d\widetilde{v}_{n} &=&\left\{ \left[ \widetilde{F}_{0}\cos \left( \varphi
_{n}\right) +\left( \widetilde{x}_{n+1}-\widetilde{x}_{n}\right) \right]
(1-\delta _{nN})\right.  \notag \\
&&-\left( \widetilde{x}_{n}-\widetilde{x}_{n-1}\right) -\alpha \widetilde{v}%
_{n}  \notag \\
&&+\lambda \left( \widetilde{x}_{n+1}-\widetilde{x}_{n}\right) ^{3}(1-\delta
_{nN})  \notag \\
&&\left. -\lambda \left( \widetilde{x}_{n}-\widetilde{x}_{n-1}\right)
^{3}+\delta _{nN}\widetilde{F}\right\} d\widetilde{t},  \label{trans} \\
d\widetilde{x}_{n}&=&\widetilde{v}_nd\widetilde{t},  \label{vel} \\
d\varphi _{n} &=&\sqrt{2\widetilde{D}_{R}}dW_n,  \label{rota}
\end{eqnarray}}
where $\widetilde{x}_{n}=x_{n}/a,\widetilde{v}_{n}=v_{n}\sqrt{M/k}/a,%
\widetilde{t}=t/\sqrt{M/k}$, $\delta _{ij}$ is the kronecker's delta, $%
\widetilde{F}_{0}=F_{0}/ka$ is the dimensionless magnitude of the imposed propulsion forces
along the random directions $\cos \left( \varphi _{n}\right) $, $W_n$ represent standard Wienner
processes with zero mean and standard deviation $\widetilde{t}$, $%
\widetilde{D}_{R}={D}_{R}\sqrt{M/k}$, with ${D}_{R}$  {as the strength of  rotational noise,} which could be caused by a vibrating motor/surface \cite{Scho} such as  in the Hexbugs-nano robots \cite{Cecilio,Paul}, $\alpha =R_{T}/\sqrt{Mk}$, is a dimensionless number which
accounts for dissipation, and where $R_{T}$ is the friction coefficient, $%
\lambda =\mu a^{2}/k$ is an introduced dimensionless parameter accounting
for nonlinearity, and $\widetilde{F}=F/ka$ is the magnitude of the imposed deformation force
along the $x-$direction, {and in dimensionless units the solids' original equilibrium length is simply given  by $L_N=N$}. { Note that Eq. (\ref{trans}) is subject to the initial conditions $\widetilde{x}_{n}(0)=%
\widetilde{v}_{n}(0)=0$ for $n=1,2...,N$, and we define $%
\widetilde{x}_{0}(t)=0$ for completeness. }
\begin{figure}[b]
\includegraphics [width=9cm]{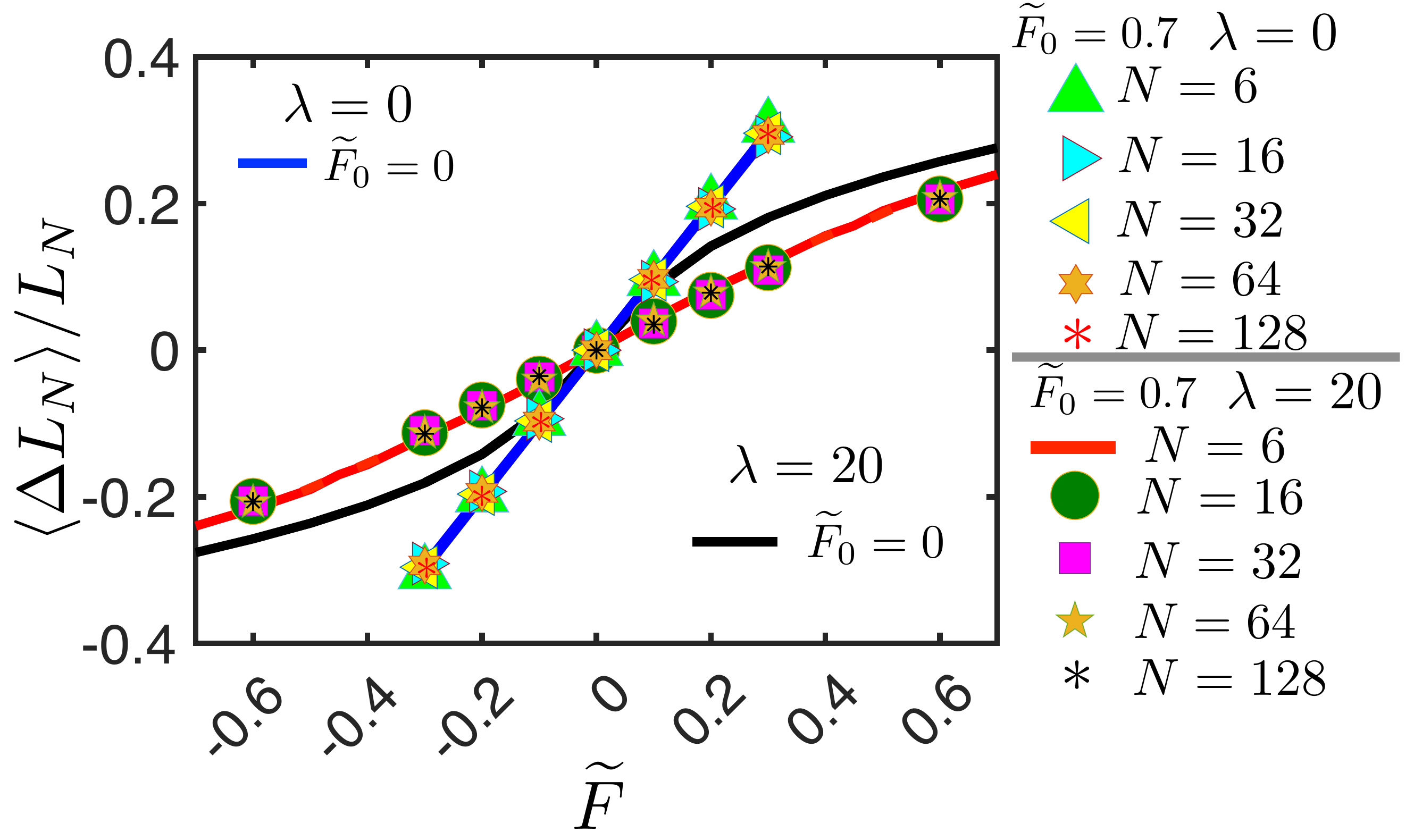} 
\caption{ { Average strain of five active solids made of $N=\{6,16,32,64,128\}$ masses (either linearly or
nonlinearly interacting) as a function of stretching and compressing forces.
The influence of activity on the deformation of the system, when nonlinearity is present, can  be
appreciated. Here, $\widetilde{D}_R=1$}. }
\label{s1}
\end{figure}
\begin{figure}
\includegraphics [width=8cm]{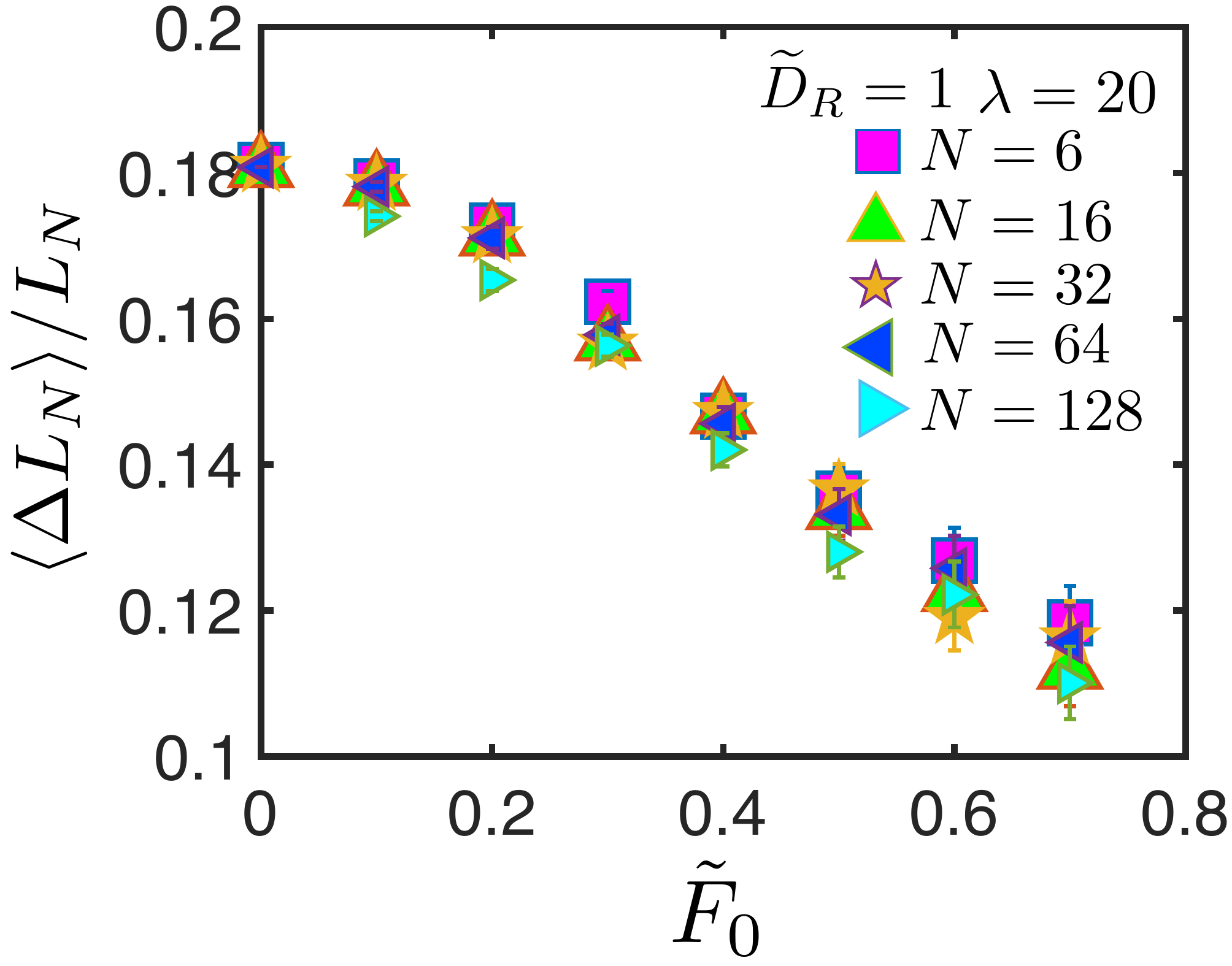}
\caption{ { Average strain of five active solids made of  $N=\{6,16,32,64,128\}$ particles, as a
function of their propulsion force. The  nonlinear spring constant is set as $%
\lambda=20$, whereas $\widetilde{D}_R=1$.} }
\label{defoU}
\end{figure}

\section{Deformation}

\label{Deformation}

{Let us start by numerically solving Eqs. (\ref{trans})-(\ref{rota})
for the case of five active solids made of $N=\{6,16,32,64,128\}$, respectively; and
look for its deformation. For the simulations, we choose a second-order
Verlet algorithm, a time step $\Delta \widetilde{t}=0.025$, a total time to
reach the system's steady state of $\widetilde{T}=\{50,300,1000, 4000,10000\}$,
and perform from  50 to 500 realizations to calculate ensemble averages. Due to the
number of interacting particles, a vectorized code had to be implemented. It
was also noticed that the larger an active solid, the longer the time it takes to
reach a steady state. A typical scenario of this steady state can be seen in
the Appendix. In addition, the following set of parameters are considered: $%
\alpha=1$, $\widetilde{F}_0=[0,0.7]$, $\widetilde{D}%
_{R}=[1,40]$, $\widetilde{F}=[-0.6,0.6]$, and $%
\lambda=\{0,5,10,20\}$. {It is worth mentioning that the previous parameters were chosen in such a way that the acting forces deform the solids in such a way that their constituent particles never overlap, that is, $\widetilde{x}_{n-1}<\widetilde{x}_{n}<\widetilde{x}_{n+1}$ for all time. The same rule will be applied thereafter.} {The results are visualized in Fig. \ref{s1}, where the
average strain ($\langle \Delta L_N \rangle/L_N$) of the mentioned  active solids versus applied force ($\widetilde{F}$) is shown.} See also supplementary material for visualization purposes. We start
validating the employed code by reproducing the linear deformation region or
Hooke's law. Here, we set $\lambda =0$, and try two propulsion forces $%
\widetilde{F}_0=\{0,0.7\}$. {The results for passive solids ($\widetilde{F}%
_0=0$) made of $N=\{6,16,32,64,128\}$ are shown as  blue solid lines in Fig. \ref{s1}. } 
\begin{figure}
\includegraphics [width=8cm]{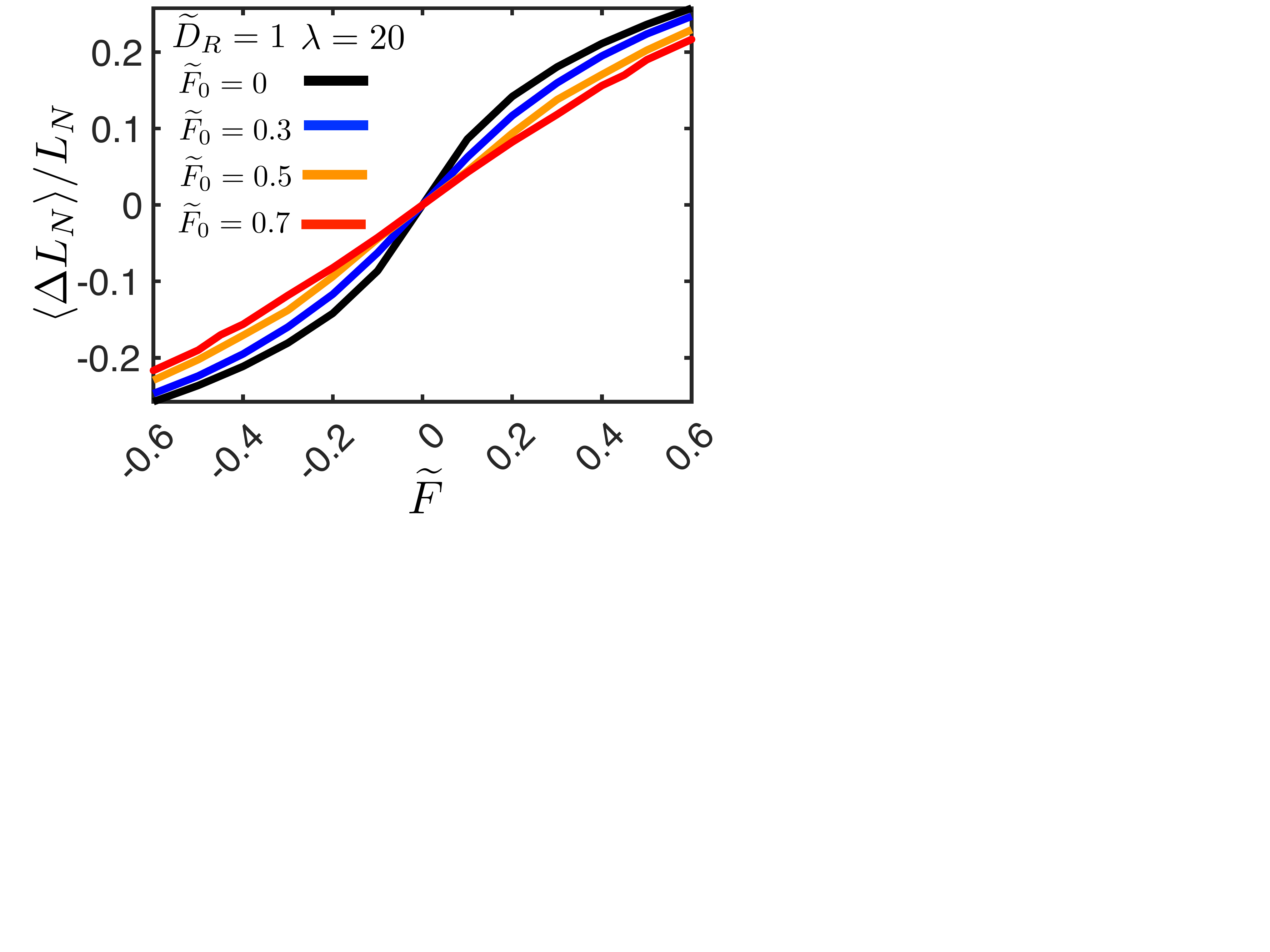}
\caption{{Average strain of an active solid made of $N=32$ particles
as a function of stress, and for several active forces. Here, the nonlinear spring constant and rotational noise are
set as $\lambda=20$ and $\widetilde{D}_R=1$, respectively. Clearly, as activity increases the active solid tends to a Hookean behavior.} }
\label{Ulabel}
\end{figure}
{{Clearly, all the solids'
deformations collapse to one curve  and the Hookean behavior is well reproduced.} {This collapsing is attributed to the fact that we are stressing the same material but of different lengths}. We now analyze some possible active forces
implications on the active solids' strain by setting $\widetilde{F}%
_0=0.7$ and $\widetilde{D}_{R}=1$, while keeping a linear interaction ($%
\lambda =0$). After performing the needed simulations, the results are
shown with the respective symbols in Fig. \ref{s1}. Interestingly, all the solids' strains fall into the same blue line, therefore it can be concluded that there is not any
activity effect on the system since both cases, $\widetilde{F}_0=\{0,0.7\}$,
have the same Hookean behavior, and of course this behavior is seen to be
independent of the solids' size. This activity-independence result for
linearly interacting masses has also been observed by Kaiser \textit{et al.} 
\cite{Kaiser} {when studying an active Rouse model. }

The next step is to add nonlinear effects by setting for
example $\lambda =20$. Thus, a first numerical experiment is to simulate the
solids' strain with null activity ($\widetilde{F}_0=0$). The results for $N=\{6,16,32,64,128\}$ are
shown in Fig. \ref{s1} as  black-solid lines. {Once again
all the passive solids' deformation fall into the same curve, thus one apparently sees only one black-solid line in Fig. \ref{s1}, but in fact there are five curves. This was done for clarity, otherwise Fig. \ref{s1} would have too much information.} These curves indicate
that nonlinear interactions have modified the solids' response to a not Hookean behavior, and 
as expected, the solids irrespective of their length, now show a
symmetric hardening effect.}

{The interesting results start here. Let us now turn on activity and keep the
nonlinearity in the solids. Consider for example $\lambda =20,%
\widetilde{D}_{R}=1$, and set activity and deformation forces, respectively,
as $\widetilde{F}_0=0.7$ and $\widetilde{F}=[-0.6,0.6%
]$. The numerical results for all the active solids are shown in Fig. \ref{s1},
where strain versus external force are represented by the symbols shown in that figure. Notice how the behavior of
the five active solids ($N=\{6,16,32,64,128\}$) is  the same and is independent of the system's size. This time, it is seen that
activity affects the systems' mechanical response, since on average, a hardening effect in the systems has
appeared. This novel observation may originate similar behaviors as the
recent experiments dealing with a hexagonal array of six active masses
colliding against a wall \cite{Martin2}, or the deformation of a structure
(made with the mentioned hexagonal array as its building block) due to the
impact of a mass through it \cite{Martin2}. Notice as well from the same figure, that the Hookean region is longer, that is, our system hardens but nearly behaves Hookean. Only after  higher stresses, one can see again the nonlinearity in the system.}
\begin{figure}[b]
\includegraphics [width=8cm]{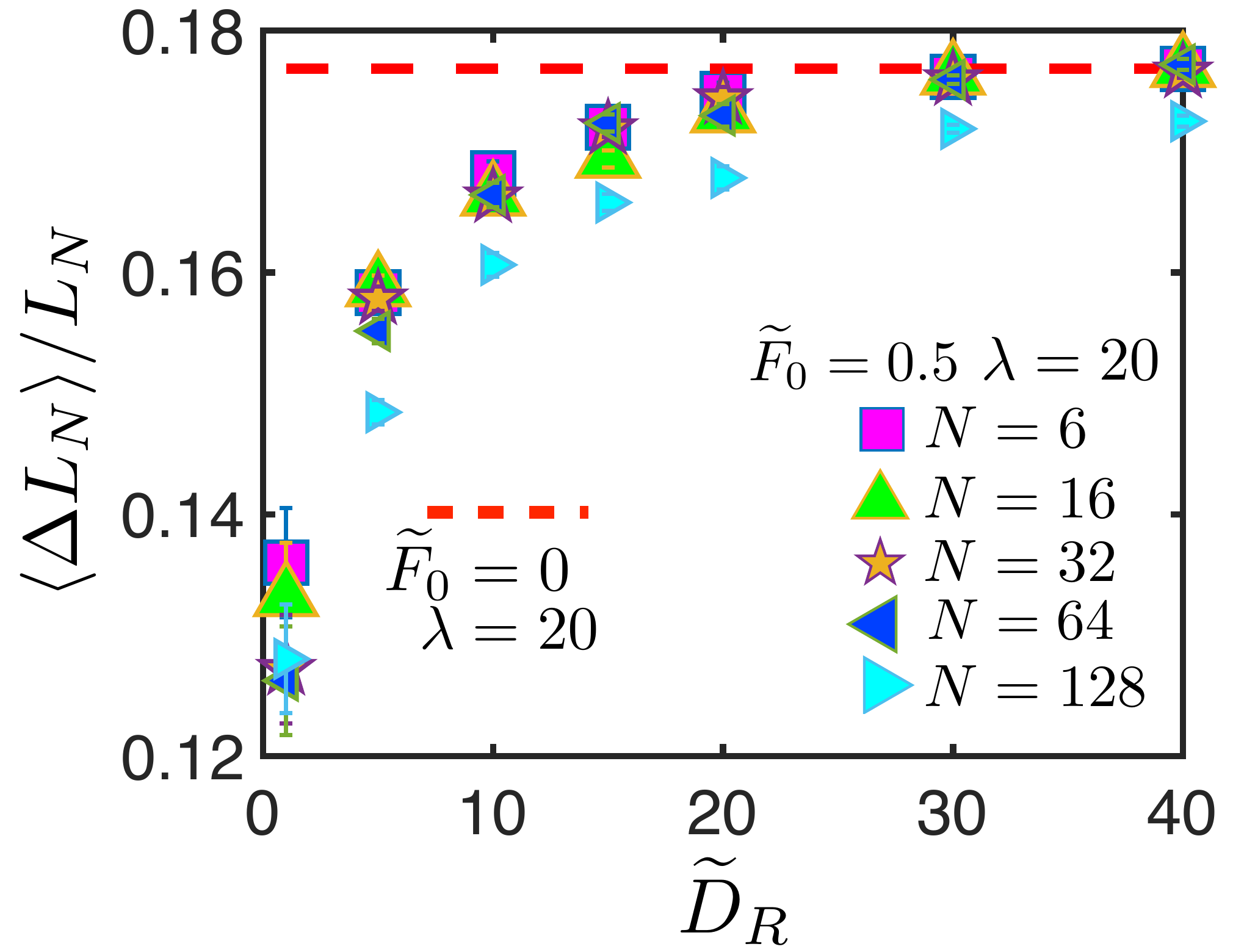}
\caption{{Average strain of four active solids made of $N=\{6,16,32,64,128\}$ particles
as a function of rotational noise. Here, the nonlinear spring constant and active force are
set as $\lambda=20$ and $\widetilde{F}_0=0.5$, respectively. For reference, the strain of a passive system is also shown as red-dashed lines.} }
\label{s4}
\end{figure}

\begin{figure*}
\includegraphics [width=17cm]{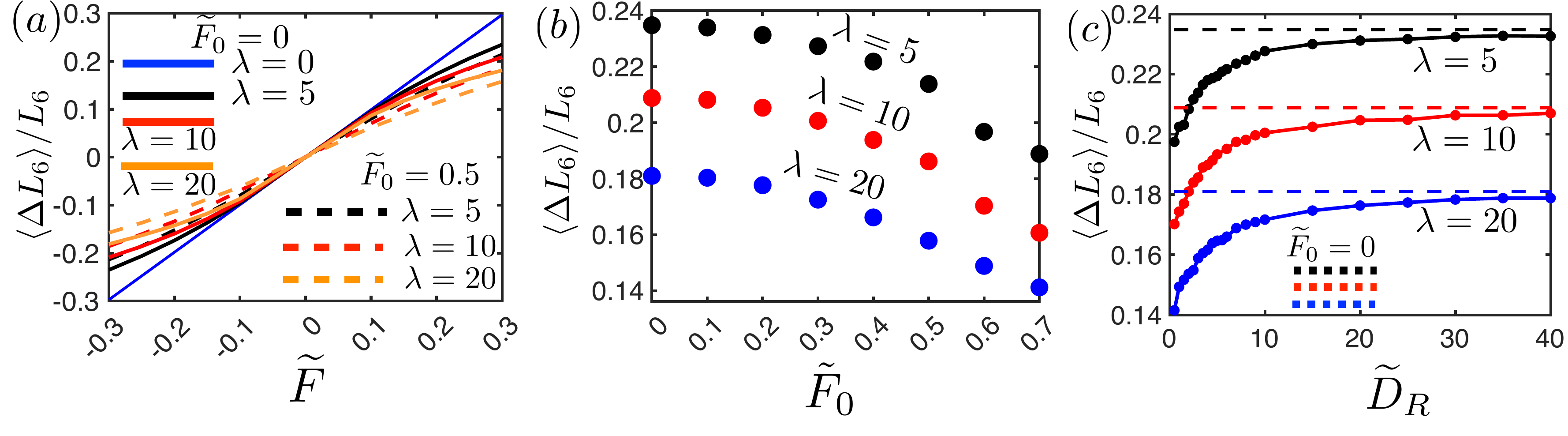}
\caption{{\ Average strain of a chain of active particles as a function of
deformation force, propulsion force, and rotational noise; and for
three nonlinear springs, namely $\protect\lambda=\{5,10,20\}$. { (a) $\widetilde{D}_R=1$. (b) $\widetilde{F}=0.3$ and $\widetilde{D}_R=1$. (c) $\widetilde{F}_0=0.5$ and $\widetilde{F}=0.3$. }  } }
\label{lambda}
\end{figure*}

{Once a self-propulsive effect has been spotted, the next step is to
understand the dependence of deformation on the magnitude of the propulsion
force. This is achieved by setting for example $\widetilde{F}=0.3$, $%
\widetilde{D}_{R}=1$, $\lambda=20$, and vary the self-propulsion force $%
\widetilde{F}_0=[0,0.7]$. Under the latter
parameters, five active solids are deformed, and their
resulting mechanical behavior (average strain vs deformation force) is
plotted in Fig. \ref{defoU}. {It can be observed that
on average the solids' deformation decreases as activity increases, and this behavior is similar for all of them  but this time size effects seem to appear. Notice for example how the pink squares ($N=6$)
are always above the cyan triangles ($N=128$). For a better accuracy, the offered measurements in Fig. \ref{defoU} also indicate their respective one standard 
deviation error. From this plot
one can also see that the maximum decrement of the strain occurs at the highest proposed propulsion force $%
\widetilde{F}_0=0.7$; however, bear in mind that higher propulsion forces may not satisfy, the no overlap condition, $\widetilde{x}_{n-1}<\widetilde{x}_{n}<\widetilde{x}_{n+1}$.} 

 {
Another way of observing activity effects in the current active solids is to  plot the solids'  average strain versus stress  for several active forces. Since this plot has already proved to be size's independent, I choose $N=32$ particles, $\widetilde{F}_0=\{0,0.3,0.5,0.7\}$, $\lambda=20$, $\widetilde{D}_R=1$, and $\widetilde{F}=[0,0.6]$. The results are shown in Fig. \ref{Ulabel}. A very interesting result from this plot is the fact that as activity increases, the strain tends to behave Hookean, that is, activity seems to dominate nonlinear effects at least for a given region. One can also notice that when the
deformation force increases, self-propulsion effects tend to
disappear. See for example the blue solid line at high stresses. Clearly, it tends to the black-solid passive behavior. This can be understood if one exaggerates the situation and
visualizes  a large stress applied to the system. This stress surely will deform the system no matter if activity is present
or not.}
{Finally, the effect of  { the strength of rotational noise}  on the deformation of the five 
active solids ($N=\{6,16,32,64,128\}$) is analyzed. By fixing the propulsion and stretching forces as $%
\widetilde{F}_0=0.5$ and $\widetilde{F}=0.3$, respectively, and varying $%
\widetilde{D}_{R}=[1,40]$ together with $\lambda=20$; we obtain Fig. \ref%
{s4}. This plot indicates that the average strain's behavior of the solids (see their respective symbols in Fig. \ref%
{s4})
is very similar, and increases as the rotational noise grows.  { For a better precision,  error bars indicating a standard deviation are also indicated.} It is also appreciated that rotational noise may be used as tune
parameter, since it enables the system to behave either as a passive solid for high 
$\widetilde{D}_{R}$ (see the horizontal red-dashed line asymptote, indicating the passive
behavior, $\tilde{F}_0=0$) or as a solid that deforms less for low $%
\widetilde{D}_{R}$. {The strain's behavior in the limit $\widetilde{D}_{R}\rightarrow 0$ is discussed in the Appendix section.  Basically, the strain seems to  asymptotically reach a minimum value as rotational noise tends to zero.}}


{Let us now analyze the effect of nonlinearity ($%
\lambda$), on the strain in our system when varying deformation force,
activity, and rotational noise.  For this experiment, the case $N=6$ will be considered
thereafter since the other solids have shown  to behave very similar.
In a  first experiment, it is set $%
\widetilde{D}_{R}=3$, $\widetilde{F}_0=\{0,0.5\}$, $\widetilde{F}%
=[-0.3,0.3]$ and $\lambda=\{5,10,20\}$. The results are
shown in Fig. \ref{lambda}(a). It can be appreciated that larger nonlinear
coefficients do not seem to change previous scenarios (see Fig. \ref{s1})
but that these coefficients only harden the system as usual. In the same
figure, the behavior of a passive solid linearly interacting and under the same conditions,
appears in solid blue lines as a reference. In a second experiment, the previous numerical parameters are also used  but this time it is fixed $\widetilde{F}=0.3
$ and vary $\widetilde{F}_0=[0,0.7]$. The
numerical results are shown in Fig. \ref{lambda}(b) which indicates that for
the three considered values of $\lambda$, the average strain in the system
behaves in the same manner. Finally, a last experiment using the same
previous parameters but fixing $\widetilde{F}=0.3$ and $\widetilde{F}_0=0.5$%
, and varying $\widetilde{D}_{R}=[0.5,40]$ is shown in Fig. \ref{lambda}(c).
Once again, nonlinearity does not change the behavior shown in Fig, \ref{s4}%
, it only hardens the system as usual. Horizontal asymptotes indicating the
respective passive strains are also shown as dashed lines. } 
\begin{figure}[b]
\includegraphics [width=8.5cm]{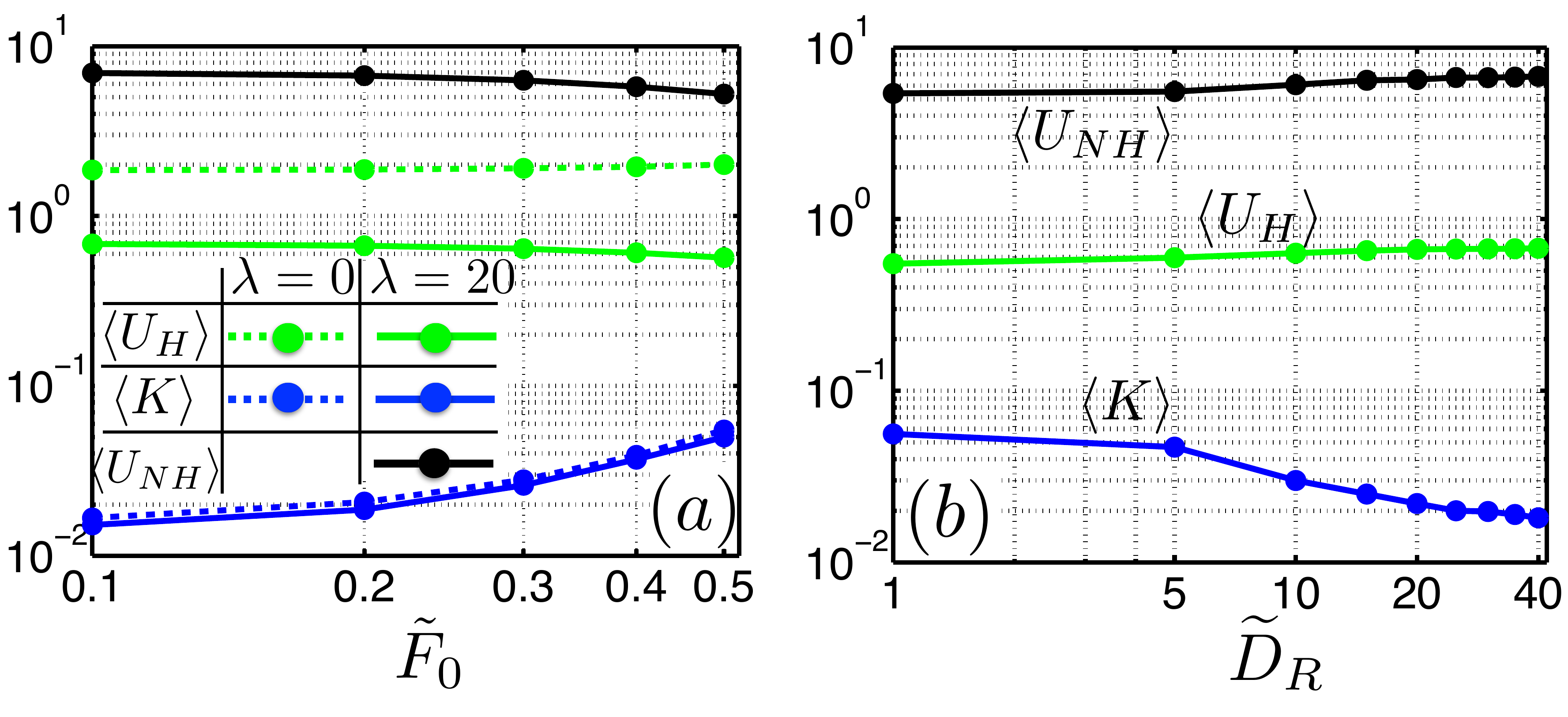}
\caption{(a) Log-Log plot of the average non-Hookean, Hookean elastic
energy, and average kinetic energy of a chain of active particles as a
function of propulsion force either linearly ($\protect\lambda =0,\widetilde{%
D}_{R}=3,\widetilde{F}=0.3$ ) or nonlinearly ($\protect\lambda =20,%
\widetilde{D}_{R}=3,\widetilde{F}=0.3$) interacting. (b) The same as in (a) but as a function of $\widetilde{D}_{R}$.}
\label{s5}
\end{figure}

\section{Energetics}
\label{Energetics}
The next goal is to understand the behavior of the dimensionless Hookean
energy ($\widetilde{U}_{H}$), non-Hookean elastic energy ($\widetilde{U}_{NH}
$), and kinetic energy ($\widetilde{K}$) as a function of propulsion force
and rotational noise in the system. These energies are defined respectively, as%
\begin{eqnarray}
\widetilde{U}_{H} &=&\frac{1}{2}\sum\limits_{n=1}^{N}\left[ -\left( 
\widetilde{x}_{n+1}-\widetilde{x}_{n}\right) ^{2}(1-\delta _{nN})\right. 
\notag \\
&&\left. +\left( \widetilde{x}_{n}-\widetilde{x}_{n-1}\right) ^{2}\right] ,
\label{UH} \\
\widetilde{U}_{NH} &=&\frac{\lambda}{4}\sum\limits_{n=1}^{N}\left[ -\left( 
\widetilde{x}_{n+1}-\widetilde{x}_{n}\right) ^{4}(1-\delta _{nN})\right. 
\notag \\
&&\left. +\left( \widetilde{x}_{n}-\widetilde{x}_{n-1}\right) ^{4}\right] ,
\label{UNH} \\
\widetilde{K} &=&\frac{1}{2}\sum\limits_{n=1}^{N}\widetilde{v}_{n}^{2}.
\label{KIN}
\end{eqnarray}
A first numerical experiment fixes $\lambda=\{0,20\}, \widetilde{D}_{R}=3$%
, $\widetilde{F}=0.3$, performs $1000$ realizations for ensemble averaging,
and vary $\widetilde{F}_0=\{0,0.1,0.2,0.3,0.4,0.5\}$. The results are shown
in Fig. \ref{s5}(a). It can be seen that for an active solid made either of nonlinear or linear
springs, their kinetic energy increases, almost identically, as a function
of the propulsion force. However, for a solid made of nonlinear springs, its
Hookean elastic energy decreases as its propulsion force increases. The
opposite occurs for a solid made of linear springs and whose elastic energy is
larger compared to the nonlinear solid. In addition, Fig. \ref{s5}(a)
illustrates that the non-Hookean elastic energy also decreases as 
propulsion in the system increases. These results suggest that as propulsion
in the active solid increases, a solid made of nonlinear springs will deform less on average compared to a system made of linear springs under the same conditions. 

The dependence of the average
non-Hookean, Hookean elastic potential energy, and average kinetic energy of
a chain of active particles ($%
\lambda=20, \widetilde{F}=0.3$) as a function of rotational diffusivity is shown in Fig. \ref%
{s5}(b). Clearly, as rotational noise grows, $\langle U_{NH} \rangle$
and $\langle U_{H} \rangle$ increase, meaning that the system is tending to
a passive behavior. This time, the kinetic
energy decreases. The decrease of energy could be originated by the fact
that for high $\widetilde{D}_{R}$, the propulsion force quickly changes
direction, thus avoiding an increase in the velocity of the system. 

\section{Rationalization}
\label{Ratio}
\begin{figure}[b]
\includegraphics [width=6.5cm]{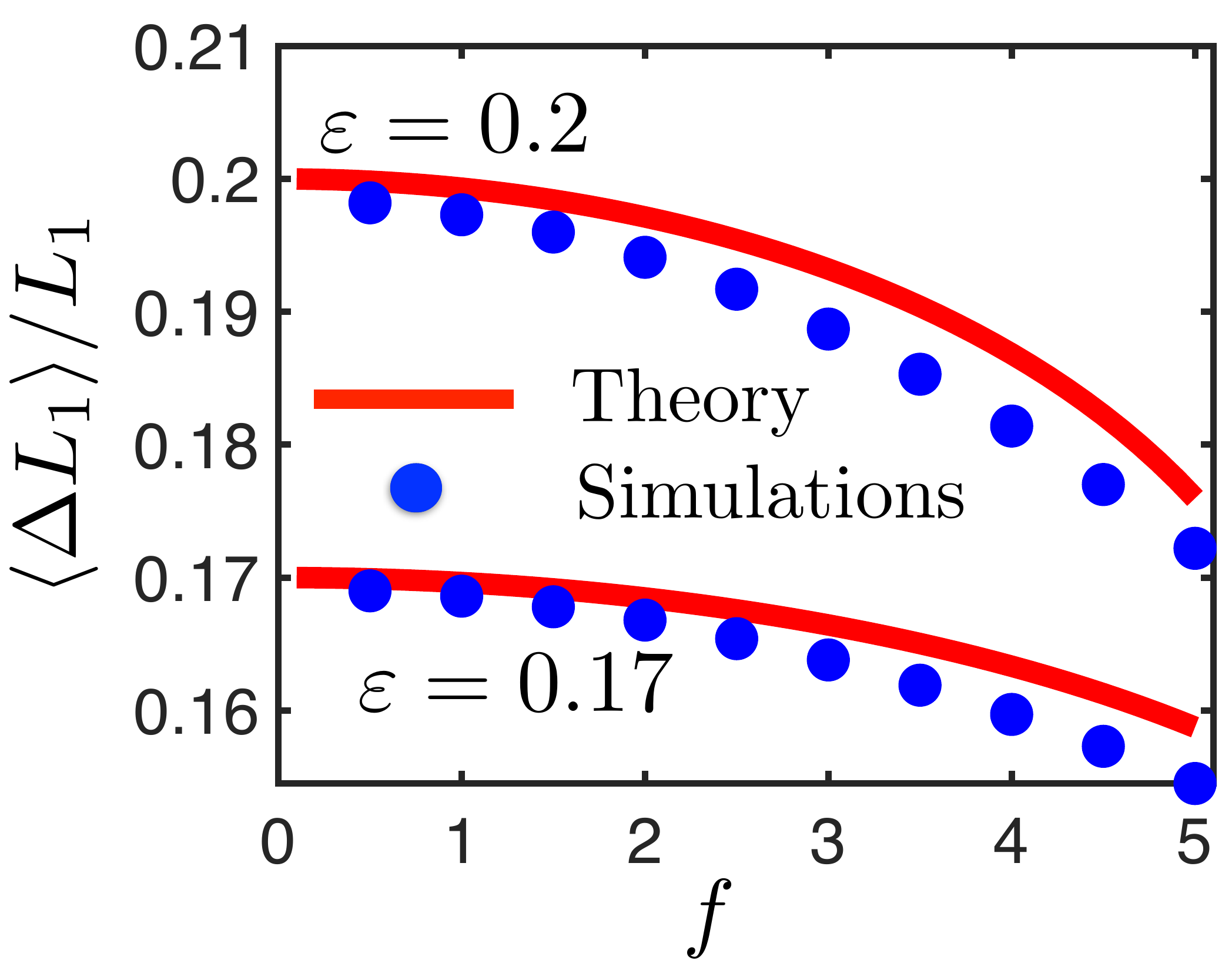}
\caption{ Comparison  between the predicted theoretical deformation (red solid lines) given by Eq. (\ref{defo}), and the numerical solution (blue circles) of Eq. (\ref%
{duffi}) with $\varepsilon =\left\{ 0.2,0.17\right\} ,\beta =2,f=\left[ 1,5%
\right] $, and $\widetilde{\Omega }=1.5$.}
\label{theo}
\end{figure}
So far we have seen the effect of rotational noise and activity on the
average deformation of an active solid.
Briefly, we have concluded that as activity in the system increases, its
average deformation decreases; and that as the rotational noise
increases, the average deformation in the system tends to the deformation of
a passive system.  {The question that in this section will be addressed is: Can we provide some theoretical explanation to the previous numerical experiments?} If we carefully analyze Eqs. (\ref{trans}) and (\ref{rota}%
), we can extract some  {sufficient } ingredients of its dynamics; namely, an active
stochastic force, a deformation force and a nonlinear interaction. Notice  that the nonlinear ingredient
was numerically observed to be crucial for the deformation of the system to become a
function of activity. Let us keep those ingredients, and analyze the dynamics of the
building block of an active solid with a minimal model. This {consists of a driven single
mass,} attached to a nonlinear spring and subject to an external constant
force. In
addition, and in order to make some theoretical progress, let us consider a weakly
nonlinear spring ($\varepsilon \widetilde{x}^{3}$) such that $\varepsilon <<1$;  with
weakly dissipation ($\alpha =\varepsilon \beta $); weakly deterministic 
driven ($\widetilde{F}_{0}=\varepsilon f\cos (\widetilde{\Omega}  \tilde{t} )$) and deformation forces ($%
\widetilde{F}=\varepsilon f_{D}$); henceforth, our building block satisfies
the so-called Duffing equation with an extra deformation force, expressly%
{
\begin{equation}
\frac{d^{2}\widetilde{x}}{{d\tilde{t}}^{2}}+\widetilde{x}+\varepsilon \beta 
\frac{d\widetilde{x}}{d\tilde{t}} {+}\varepsilon \widetilde{x}^{3}=\varepsilon
f\cos (\widetilde{\Omega}  \tilde{t} ) +\varepsilon f_{D}\mathbf{,}
\label{duffi}
\end{equation}%
where the dimensionless angular frequency of the driven force ($\widetilde{\Omega }=\Omega/\sqrt{k/M}$) near
resonance satisfies $\widetilde{\Omega }=1+\varepsilon \omega$, with $\omega$ a driver dimensionless angular  frequency of order one.  {Note that the justification for the implementation of a deterministic driven force in this model, is given in Section \ref{expe}.} Following the method of strained coordinates, a slow variable, $\tau =\varepsilon \tilde{t}$, will also be introduced.} Seeking a
perturbation solution to Eq. (\ref{duffi}) of the form $\widetilde{x}(\tilde{t},\tau%
)=\widetilde{x}_{0}(\tilde{t},\tau)+\varepsilon \widetilde{x}_{1}(\tilde{t},\tau)+\varepsilon
^{2}\widetilde{x}_{2}(\tilde{t},\tau)+O(\varepsilon ^{3})$ and substituting it into Eq. (\ref{duffi}), it is possible to prove (after neglecting terms
of $O(\varepsilon ^{3}),$ and defining the operators $L[]=\partial ^{2}/\partial \tilde{t}^{2} +1$ and $P[]=-2\partial ^{2}/\partial \tilde{t}\partial \tau- \beta\partial/\partial \tilde{t}$) that the following coupled system arises %
\begin{eqnarray}
L[\widetilde{x}_{0}] &=&0,  \label{1} \\
L[\widetilde{x}_{1}] &=& P[ \widetilde{x}_{0}]%
-\widetilde{x}_{0}^{3}+f\cos \left( \tilde{t}+\omega \tau\right) +f_{D},  \label{2} \\
L[\widetilde{x}_{2}] &=&P[ \widetilde{x}_{1}]-\frac{\partial ^{2}\widetilde{x}_{0}}{\partial 
\tau^{2}}-\beta \frac{\partial \widetilde{x}_{0}}{\partial \tau}-3\widetilde{x}_{0}^{2}\widetilde{x}_{1}.  \label{3}
\end{eqnarray}
After eliminating secular
terms in Eqs. (\ref{1})-(\ref{3}), their remaining inhomogeneous terms will be harmonic, subharmonics, and constants terms. The
constant terms precisely provide the information for the net deformation in the
system as a function of stretching/compressing forces. This deformation can  be explicitly calculated, namely%
\begin{equation}
\Delta L=\varepsilon f_{D}-\varepsilon ^{2}\frac{3A^{2}f_{D}}{2},
\label{defo}
\end{equation}%
where $A$ represents the steady amplitude of Eq. (\ref{duffi}). After  the elimination of the secular terms, this amplitude is found to satisfy \cite{Kevo}
\begin{equation}
\beta ^{2}A^{2}+\left( 2\omega A-\frac{3A^{3}}{4}\right) ^{2}=f^{2}.
\label{rc}
\end{equation}%
The latter expression couples the steady amplitude and the propulsion force. Thus Eq. (\ref{defo}) indicates that only to $O(\varepsilon ^{2})$ and due to nonlinearity, activity will influence the system's deformation. To prove that our analytical expression, Eq. (\ref{defo}), correctly
describes such a deformation in our building block; we numerically solve Eq. (\ref%
{duffi}) under the following parameters: $\varepsilon =\left\{ 0.2,0.17\right\} ,\beta =2,f=\left[ 1,5%
\right] $, and $\widetilde{\Omega }=1.5$. {Note that the previous parameters were chosen in such a way that the acting forces in the building block  do not deform it beyond a third of its original equilibrium distance $a$.} The results can be seen in Fig. \ref{theo}, where the deformation in the building block as a  function  of self-propulsion is illustrated. The predicted theoretical deformation using Eq. (\ref{defo}) is shown as a  red line, whereas the numerical solution of Eq. (\ref{duffi}) is shown in blue circles. This figure shows an excellent agreement between theory and simulations, hence one can argue that Eq. (\ref{defo}) contains  {sufficient }ingredients to understand the dependence of strain on activity  in our building block. In addition, one can see that this figure possesses a very similar behavior as Fig. \ref{defoU} and Fig. \ref{lambda}(b), which allow us to have some physical insight on the behavior of those plots.


\section{Possible experiment}
\label{expe}
{At this stage, theoretical and numerical analysis have been offered but, Is it possible to build an experiment? What are the minimum components to make an active solid? To explicitly see this, let us  take some macroscopic experimental values reported by Lowen {\it et. al} \cite{Scho}. Their experiment considers a vibrobot which as it will be seen, may be a candidate for experimentally realizing our active solid. Using  a specific set of their data for a given vibrobot, namely, $M = 0.004 kg$, $U=0.09m/s$ (propulsion speed), $\Omega=0.7s^{-1}$ (intrinsic angular velocity), $D_B=8\times10^{-5}m^2/s$, $D_R=2.59s^{-1}$, and $R_T/M=6.46s^{-1}$, and assuming a strong trap constant (see \cite{Cecilio} for the strong trap condition) of $k= 0.1 kgs^{-2}$, a deformation force  $f_D=1N$, and a separation equilibrium distance between robots of $a=0.15m$; leads to the first row of  Table \ref{T1}, where orders of magnitude for some relevant experimental parameters are indicated.}
 \begin{table}[h!]
\centering
\begin{tabular}{||c c c c c||} 
 \hline
 $$ & $\alpha$ & $\widetilde{D}_R $ & $\widetilde{\Omega} $  & $\widetilde{F}_0$  \\ [0.5ex] 
 \hline\hline
 Vibrobot & 1.4  & 0.54 & 0.14  & 0.17 \\ 
 \hline
 Hexbug & 1.9  & 0.02 & 0.17  & 0.14 \\ [1ex] 
 \hline
\end{tabular}
\caption{{Some relevant experimental dimensionless quantities, for a vibrobot \cite{Scho}  and a hexbug-nano robot \cite{Cecilio}, appearing in Eqs. (\ref{trans}) and (\ref{rotahex}).}}
\label{T1}
\end{table}}
{In addition, Sandoval {\it et. al} \cite{Cecilio} considered the dynamics of trapped hexbug-nano robots (another possible candidate for the experimental realization of the current system). They emphasized that hexbugs also posses an intrinsic torque which makes those robots to rotate with an angular velocity of around $\Omega=1.9s^{-1}$. Thus by using the parameters describing those robots namely, $U=0.13m/s$ (propulsion speed), $M = 0.0072 kg$, and $S_T=0.0027$ (Stokes translational number), where $S_T= \tau_M/\tau_R$ with $\tau_M = M/R_T$, and $\tau_R = 1/D_R$; and assuming a strong trap constant of $k= 0.9 kgs^{-2}$ together with $D_B=8\times10^{-5}m^2/s$, $a=0.15m$, and $f_D=1N$; one gets the second row of Table \ref{T1}. Interestingly, this table shows that the dimensionless experimental quantities fall into the numerical parameters used during the simulations. This directly establises a connection between theory, simulations and experiments. Notice as well that both of the mentioned teams propose an internal degree of freedom ($\varphi$) whose dynamics in dimensionless variables reads
  \begin{equation}
d\varphi =\tilde{\Omega}d\widetilde{t}+\sqrt{2\widetilde{D}_{R}}dW,  \label{rotahex}
\end{equation}
where $W$ represents a standard Wienner process. Therefore, based on Table \ref{T1}, one could approximate the rotational dynamics of a hexbug-nano simply as $d\varphi =\tilde{\Omega}d\widetilde{t}$, hence the projected active force, $\varepsilon fcos(\varphi)$, would become $\varepsilon fcos(\tilde{\Omega})$, thus justifying the proposed driven force in Eq. (\ref{duffi}). Moreover, the discussed order of magnitude  for hexbugs can be visualized in the supplementary material video: Home-made-experiment.mov, where two hexbugs are practically  rotating  clockwise and with a very small angular noise. A snapshot of the mentioned video is offered in 
Fig. \ref{exp}. Here, springs taken from pens, two glasses as corrals, two-hexbugs-nano, two walls, and a pulling force (hand of the author) are shown to conform an active solid deformation experiment. These would be the minimum components for a more sophisticated experiment.}
\begin{figure}
\includegraphics [width=8.5cm]{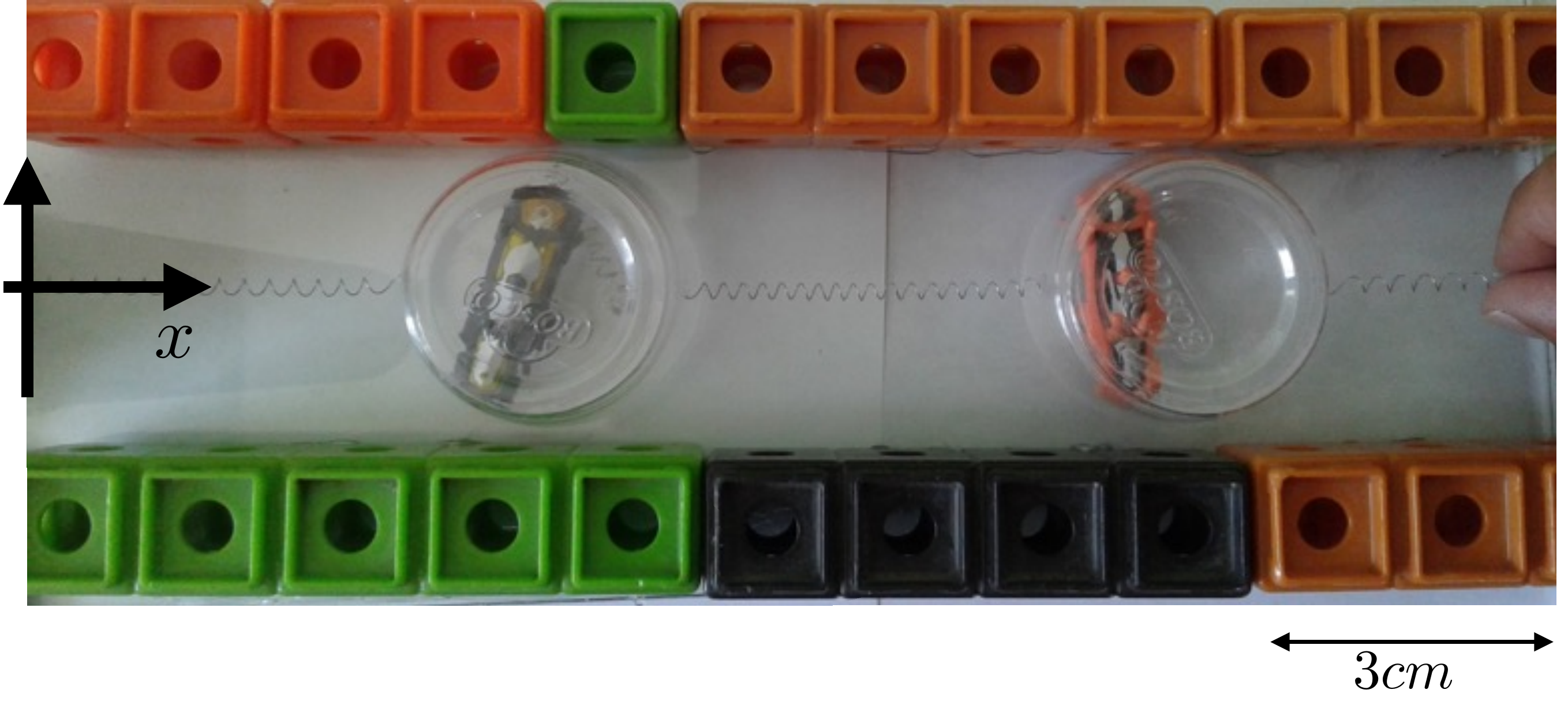}
\caption{ { Possible experimental realization of an 1D active solid using hexbugs-nano robots. Notice that although the robots freely spin and hence propel towards any direction, only their active forces projections (and due to the presence of walls) along the $x$-axis affect the active solid dynamics. }}
\label{exp}
\end{figure}

\section{Conclusions}
\label{summa}
It has been shown that a mixture of self-propulsion and nonlinear interactions; in particular, a hard spring symmetric potential, is  {sufficient} to render an active solid the ability to deform less  on average --compared to a passive one-- when subject to a deformation force. The latter observation may serve as another mechanism to creating novel active metamaterials (a material with its own energy, that given its  similarity with the movie 'Flubber' \cite{flubber}, it could be called a flubber material). The dependence of the studied solids' strain on activity was explained by introducing a minimal model that proved that nonlinearity is the responsible for this feature. The strain behavior versus stress from all the analyzed active solids collapsed to a single curve, meaning that the properties of our new material remain the same irrespective of the length of the sample. {It was also discovered --see Fig. \ref{Ulabel}-- that as an active solid under stress gains internal energy, it tends to behave Hookean, that is, a linear behavior seems to dominate over a nonlinear one. This switching between a Hookean and a nonlinear behavior has not been observed before and may be related to the way adherent cells modify their elastic properties according to their needs \cite{Dogic}. In addition, by using available experimental data \cite{Scho,Cecilio}, it was possible to show that the numerical parametes used in this work fall into real orders of magnitude of an experiment, thus establishing a connection between theory, simulations and experiments.}

{Although further investigations are needed to see if this 'flubber material' has some practical applications, probably, those applications may be in a close manner to classical metamaterials, which have potential impact on energy absorbing devices (helmets, footwear), noise reduction, automobile parts, prosthetic implants, among others. Future directions, and already in progress, will be to evaluate the effect of nonlinearity and activity, on the speed of waves propagating through this system, as well as heat transferring on these structures. I believe that the present active solid can be experimentally realized and its extension to higher dimensions, already in progress by the author, may be significative for the physics and materials science communities.}


\section{Data availability statement}
This manuscript does not have associated data

\section{Acknowledgements}

M. S. thanks Consejo Nacional de Ciencia y Tecnologia, CONACyT for support. The author also thanks nature for giving him extra time.
\begin{figure}
\includegraphics [width=6cm]{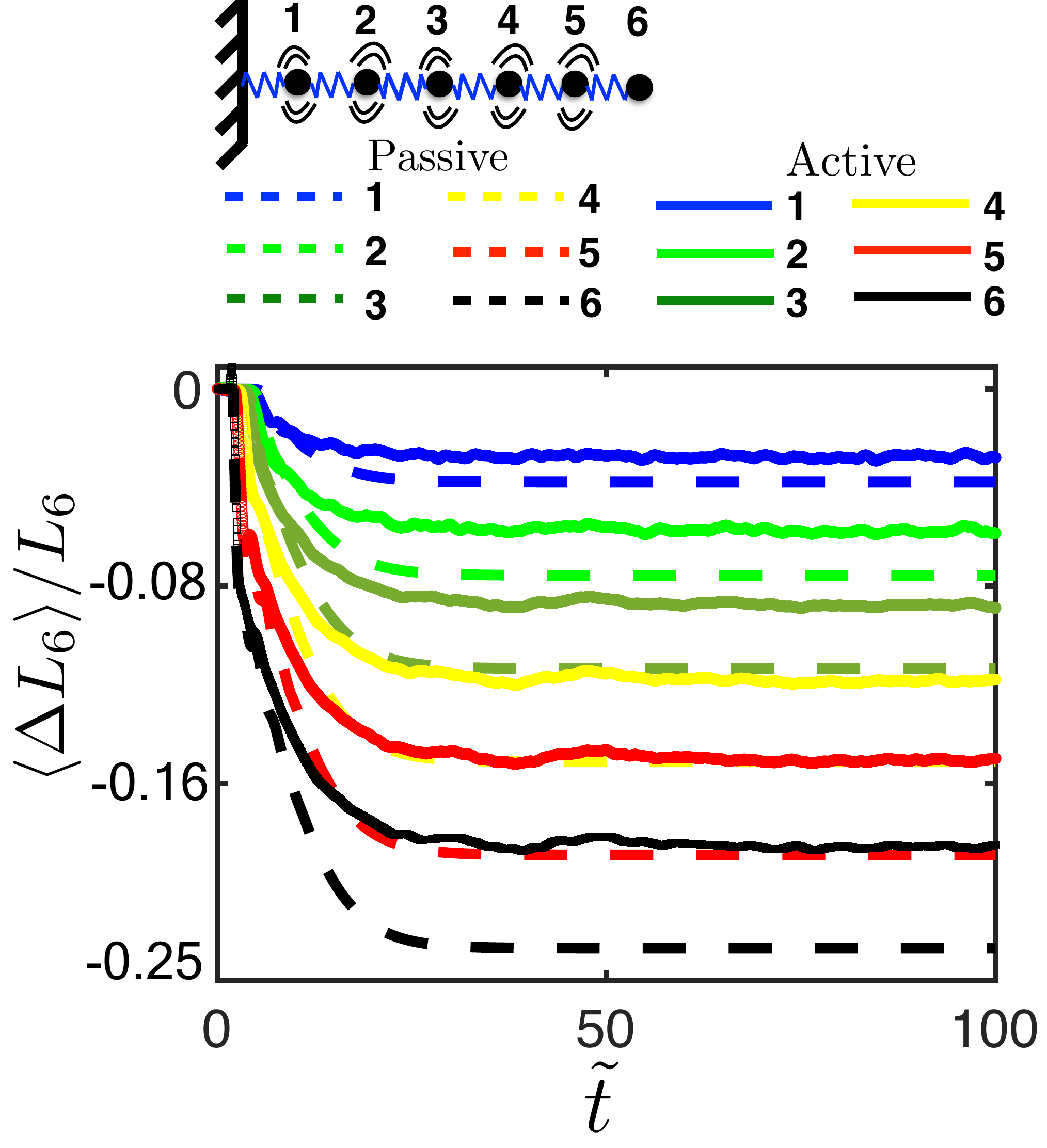}
\caption{{ Strain of each particle in an active/passive solid made of  $N=6$ masses reaching a steady state.}}
\label{A1}
\end{figure}

\section{Appendix}
\subsection{Reaching a steady state}
{In this appendix, the way  the strain of each particle, composing either an active or passive solid made of  $N=6$ masses, reaches a steady state is illustrated in Fig. \ref{A1}. The numerical values for the simulations are $%
\lambda=20, \widetilde{D}_{R}=3, \widetilde{F}=3, \widetilde{F}_0=0.5$, $\Delta \widetilde{t}%
=0.02$, a total time of $\widetilde{T}=100$, and $1000$ realizations
to calculate the respective ensemble averages.  Dashed lines represent the passive chain dynamics, whereas solid lines represent the active counterpart. Clearly, self-propulsion decreases on average the strain in the system.}
\subsection{Case $\widetilde{D}_R \rightarrow 0$}
 {Figure \ref{A2} illustrates the convergence to a minimum deformation of an active solid made of $N=16$ particles, as  rotational noise tends to zero ($\widetilde{D}_R \rightarrow 0$). Here, the domain of the rotational diffusivity varies along three orders of magnitude, namely,  $\widetilde{D}_R=[0.005,0.05,0.5,1]$. The parameters for the simulations are 500 realizations, $\widetilde{F}_0=0.5$, $\widetilde{F}=0.3$, and $T=2000$. Note that the case $\widetilde{D}_R=0$ does not provide an absolute  minimum since our model assigns random angles to the active particles, thus depending on the random distribution of active forces, the active solid may deform more or less from the true minimum due to activity. Some trivial configurations could also be proposed. For example, all the active forces having $\pi$ radians. This leads to a minimum deformation of the active solid when subject to a stretching force. However, this is not a symmetric configuration when subject to a compressing force. Other simpler configuration that has seen very effective for a hardening effect, is to locate at each particle a 1D central force field \cite{Dogic}. Clearly, this force configuration will stiffen the solid.}
\begin{figure}
\includegraphics [width=6cm]{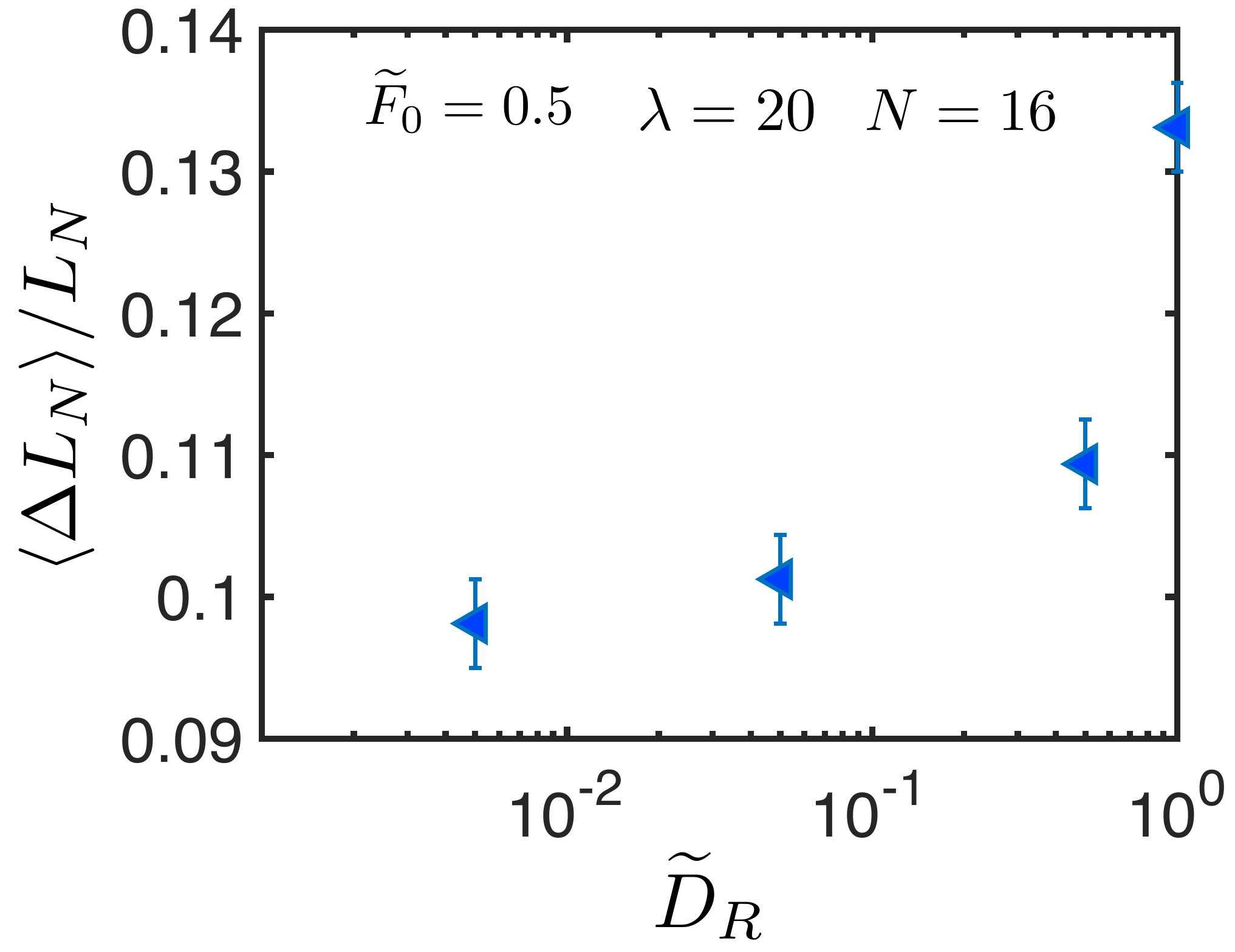}
\caption{{Convergence to a minimum deformation of an active solid made of $N=16$ particles, as  rotational noise tends to zero. Its domain in this plot is  $\widetilde{D}_R=[0.005,0.05,0.5,1]$.}}
\label{A2}
\end{figure}

\bibliography{flub}

\end{document}